# Evidential Analysis: An Alternative to Hypothesis Testing in Normal Linear Models

Brian Dennis, Department of Fish and Wildlife Sciences and Department of Mathematics and Statistical Science, University of Idaho, Moscow, Idaho, USA.

Mark L. Taper, Department of Ecology, Montana State University, Bozeman, Montana, USA.

José M. Ponciano, Department of Biology, University of Florida, Gainesville, Florida 32611, USA

Abstract. Statistical hypothesis testing, as formalized by 20th Century statisticians and taught in college statistics courses, has been a cornerstone of 100 years of scientific progress. Nevertheless, the methodology is increasingly questioned in many scientific disciplines. We demonstrate in this paper how many of the worrisome aspects of statistical hypothesis testing can be ameliorated with concepts and methods from evidential analysis. The model family we treat is the familiar normal linear model with fixed effects, embracing multiple regression and analysis of variance, a warhorse of everyday science in labs and field stations. Questions about study design, the applicability of the null hypothesis, the effect size, error probabilities, evidence strength, and model misspecification become more naturally housed in an evidential setting. We provide a completely worked example featuring a 2-way analysis of variance.

Key words: evidence, evidence functions, linear models, Neyman-Pearson, hypothesis testing, Kullback-Leibler, Schwarz information criterion, SIC, BIC, AIC, noncentral distribution.

## 1. Introduction

In this paper we construct an evidential-assessment approach to classical statistical analyses based on the univariate normal linear model with fixed effects. The model family includes the standard models for one and two sample t-tests, simple linear regression, multiple regression, analysis of variance (ANOVA) under various experimental designs, and models with mixed categorical and quantitative predictor variables. These models form much of the parade of statistical methods in the usual graduate course on applied statistics for researchers. The analysis scenarios in normal linear models are ordinarily handled by the machinery of Neyman-Pearson (NP) hypothesis tests and accompanying confidence intervals [1]. While the inferences in NP testing have many desirable statistical properties, they are weighed down by some often-discussed drawbacks, namely, the fixed Type 1 error rate ($\alpha$) not being dependent on sample size, ambiguity concerning what constitutes evidence for the null hypothesis, controversies over the meaning of $p$-values, and the possible distortion of results caused by model misspecification [1]. The evidential approach mitigates these drawbacks of hypothesis testing.

The evidential approach to statistical inference was developed in large part by Royall [2] based on earlier proposals (e.g. [3]), but the methodology was confined to models without unknown parameters. Lele [4] and Taper and Lele [5] expanded the definition and understanding of “evidence functions” in evidential statistics. Taper and Ponciano [6] described and compared inference concepts under evidential analysis, frequentist analysis, and Bayesian analysis. Comparative properties of evidence functions and NP hypothesis testing under model misspecification were studied by Dennis et al. [7]. Estimation of different levels of uncertainty in evidential analysis under model misspecification was described by Taper et al. [8]. Cahusac [9] provided a comprehensive account of how Royall's [2] original evidence concepts can be implemented for standard statistical analyses.

Royall's [2] original evidence concepts were cast only for models with no unknown parameters (so-called simple statistical hypotheses). More recently, evidential analysis is being extended to models with unknown parameters (also known as composite hypotheses). Specifically, evidence functions for composite models can be constructed from a class of "information theoretic" model selection indices [7, 8].

The evidential approach is implemented in the form of an evidence function: a statistic for comparing two models by estimating, based on data, the difference of their divergences from the data generating process, i.e. truth [4]. In a leading formulation, an evidence function is a difference of penalized maximized log-likelihoods and is essentially a contrast between two generalized entropy discrepancies [7, 8]. A consequence of this definition is the salient property that the probabilities of misleading evidence, error probabilities analogous to Type 1 and Type 2 errors in hypothesis testing, *both* approach 0 as sample size increases. Furthermore, the conclusions of an evidential analysis retain some robustness to model misspecification [7], while the uncertainty inherent in an evidential analysis can be assessed under the general assumption of model misspecification [8]. Thus, the evidential approach can remedy some shortcomings of NP hypothesis testing.

Here we show that the concepts of evidential analysis are ready-made for, and easily folded into, the existing hypothesis-testing framework of normal linear statistical models. To emphasize this, our notation and development cleaves as much as possible to the standard introductory treatment of linear model theory (for example, [10, 11]). Our presentation is intended to be accessible to data analysts who are familiar with the matrix formulation of linear models. We provide a completely worked example of evidential analysis for a common statistical problem (2-way ANOVA) as presented in introductory statistics courses. The example ordinarily would be handled with textbook NP hypothesis testing. We demonstrate how study design can be based in evidential analysis on the probabilities of misleading evidence. The overall approach can be adapted to many other statistical models and scenarios for which power calculations are feasible.

## 2. The structure of evidential analysis

The evidential approach to statistical inference begins with an *evidence function*. Two probability models, with respective pdf's denoted by $f_1(y,\theta_1)$ and $f_2(y,\theta_2)$, are under contention as models of the probabilistic process generating observations $y_1, y_2, \cdots, y_n$. Here $\theta_1$ and $\theta_2$ are parameter vectors. The thematic goal of an evidential analysis for models $f_1$ and $f_2$ is to make a statistical inference about which model more accurately resembles the probability mechanism that generated the data. The implied model quality is measured by some quantity defining the divergence of a model $f$ from the true data generating mechanism $g$. Here we adopt the Kullback-Leibler (KL) divergence measure, given by

$$K(g,f) \equiv \mathrm{E}_g\left[\log\left(\frac{g(Y)}{f(Y)}\right)\right] = \sum g(y)\log\left(\frac{g(y)}{f(y)}\right), \tag{1}$$

that is, the expected value of $\log[g(Y)/f(Y)]$ with respect to pdf $g$, as the basis for the methods presented in this paper. The KL divergence is also known as the cross-entropy or the relative-entropy [12]. The two probability distributions indicated by $f$ and $g$ could be discrete, continuous, or mixed discrete/continuous (i.e. the expectation is a sum, integral, or both), but they both must give positive probabilities to the same sample outcomes. The KL divergence underlies much of maximum likelihood estimation theory for standard statistical methods [13, 14]. Other divergence measures such as the Hellinger distance [15, 16, 4] can be used to form evidence functions having different statistical properties better suited to different purposes such as decreased sensitivity to outliers.

An evidence function is a statistic that confers upon the evidential analysis certain desirable statistical properties. A full list of properties is enumerated elsewhere [6]; for the current discussion, the most relevant is that the probability of picking the right model under the evidential decision rules must approach 1 as the sample size $n$ increases.

The evidence function we use is built on the following concepts (for a more detailed treatment see [8]). Define $\Delta K$ to be the difference of KL divergences of approximating models $f_1$ and $f_2$ from $g$ (the data-generating process), using the versions of $f_1$ and $f_2$ that are "closest" to $g$. An evidence function, when divided by $n$, is a consistent statistical estimator of $\Delta K$, that is, an estimator of which model is closest to the data generating mechanism $g$ [4, 8]. The explicit definition of $\Delta K$ and associated quantities that we adopt in this paper are as follows:

$$\Delta K = K(g, f_1^*) - K(g, f_2^*) \ . \qquad (2)$$

Here $f_j^* = f_j(x, \theta_j^*)$, where $\theta_j^*$ is the value of the parameter vector $\theta_j$ that minimizes the KL divergence of $f_j(x, \theta_j)$ from $g(x)$, the best version of $f_j$ for representing truth under the KL criterion. If $f_1$ is the better model, then $\Delta K < 0$, and if $f_2$ is the better model, then $\Delta K > 0$. If $f_1^*$ is closest to $g$, and $f_1$ is nested within $f_2$ (i.e. the parameter space for $f_1$ is nested within that of $f_2$, in the classical null/alternative NP hypothesis setup), or $f_1$ and $f_2$ are overlapping with $f_1^*$ in the region of overlap, then $\Delta K = 0$, because $f_1^*$ and $f_2^*$ are then the same model.

Because the parameters in the vectors $\theta_1^*$ and $\theta_2^*$ are unknown, they must be estimated. Maximum likelihood estimation provides statistically consistent estimates of $\theta_1^*$ and $\theta_2^*$. The likelihood function for the observations $y_1, y_2, \cdots, y_n$ under model $f_j$ is

$$L_j(\theta_j) = \prod_{i=1}^{n} f_j(y_i, \theta_j) \, . \tag{3}$$

The maximum likelihood (ML) estimate $\hat{\theta}_j$ is the vector of parameter values that jointly maximize $L_j(\theta_j)$. The ML estimate is known to converge in probability to $\theta_j^*$ [17].

A convenient evidence function based on KL divergence is the difference of Schwarz Information Criteria (SIC's, also known as BIC's; [18]):

$$\Delta\mathrm{SIC} = \mathrm{SIC}_1 - \mathrm{SIC}_2, \tag{4}$$

where

$$\mathrm{SIC}_j = -2\log\left[L_j(\hat{\theta}_j)\right] + r_j \log(n) \, , \tag{5}$$

in which $L_j(\hat{\theta}_j)$ is the maximized likelihood function for model $f_j$ $(j = 1, 2)$, and $r_j$ is the number of parameters estimated in $\theta_j$. If $\Delta\mathrm{SIC} > 0$, model 2 is estimated to be closer to $g$ than model 1, while if $\Delta\mathrm{SIC} < 0$, model 1 is estimated to be closer. [1]

One of the advantages of using an evidence function based on SIC is that it is related to Wilks' [19] generalized likelihood ratio test statistic for NP hypothesis testing when one or both models have unknown parameters. In NP hypothesis testing when models have unknown parameters, ordinarily one of the models (the null hypothesis) is nested within the other (the alternative hypothesis). Specifically, model 1 is formed from model 2 by imposing one or more restrictions on parameter values, often by fixing their values equal to known constants. Then the vector $\theta_1$ contains only the parameters from $\theta_2$ that remain unrestricted and unknown. The generalized likelihood ratio statistic is

$$G^2 = -2\log\left[\frac{L_1(\hat{\theta}_1)}{L_2(\hat{\theta}_2)}\right], \tag{6}$$

where $L_2$ is the likelihood function for the full model, and $L_1$ is the same likelihood except it is evaluated at the restricted parameter values and maximized over the remaining unknown parameters. It can be seen that

$$\Delta\mathrm{SIC} = G^2 - \nu\log(n), \tag{7}$$

where $\nu = r_2 - r_1$. The consequence is that the well-studied distribution theory for $G^2$ can be commandeered for use in evidential approaches.

[1] If models 1 and 2 were two of multiple models under consideration, one can denote each pairwise ΔSIC value with two subscripts. In the convention proposed by Taper et al. [8], the evidence function in Eq. (7) would be written as $\Delta\mathrm{SIC}_{2,1}$. That is to say, a positive ΔSIC is evidence for the model indicated by the first subscript over the model indicated by the second subscript. Here we are dealing solely with the NP setup of two models, one nested within the other, and we dispense with the subscripts to reduce clutter.

An evidential analysis using an evidence function such as ΔSIC picks two threshold values, $k_1$ and $k_2$ ($k_1 < 0 < k_2$) that produce a trichotomy of outcomes [2, 7]. An evidential analysis deems there is strong evidence for model $f_1$ if $\Delta\mathrm{SIC} < k_1$, strong evidence for model $f_2$ if $k_2 < \Delta\mathrm{SIC}$, and weak or inconclusive evidence if $k_1 < \Delta\mathrm{SIC} < k_2$. Taper et al. [8] proposed four threshold $k$ values giving five classifications (strong evidence for model 1, prognostic evidence for model 1, weak evidence, prognostic evidence for model 2, strong evidence for model 2) for the point value of the evidence function, in order to provide investigators with more descriptive outcomes. Additional $k$ values are readily added to an analysis using the methods described in this paper, so for brevity our discussions here concentrate on just specifying $k_1$ and $k_2$.

As mentioned above, an evidence function endows the analysis with desirable frequentist error properties. In particular, the two probabilities of misleading evidence given by

$M_1 = P(k_2 < \Delta\text{SIC} \mid \text{model } f_1 \text{ generated the data})$ (8)

and

$M_2 = P(\Delta\text{SIC} < k_1 \mid \text{model } f_2 \text{ generated the data})$ (9)

asymptotically approach 0 as sample size $n$ increases [2, 7]. We note that when $f_1$ is nested in $f_2$, $M_1$ does not go to 0 but rather approaches a positive constant value if ΔAIC, the difference of AIC values, is used as an evidence function [7]. Thus strictly speaking ΔAIC is not an evidence function but rather has properties more akin to NP hypothesis testing. Accompanying these two error probabilities are two probabilities of weak or inconclusive evidence, usually denoted $W_1$ and $W_2$, corresponding to the event $k_1 < \Delta\text{SIC} < k_2$ under models 1 and 2 respectively, and they both approach zero as sample size increases. The probabilities $V_1$ and $V_2$ of strong, correct evidence for model $j$ ($j = 1, 2$), given model $j$ generated the data, become

$V_j = 1 - (W_j + M_j)$ . (10)

If model $j$ generated the data, $V_j$ is monotonically increasing and approaches 1 as sample size increases [4, 7]. Here V stands for “veridical” or truth-like.

The choice of information index (SIC, HIC, etc. [7, 8]) on which to base an evidence function has inferential consequences. Using ΔSIC (Eq. _ 7_) will weight the inference toward simpler models and might be chosen if the investigation is averse to including spurious predictor variables (or other model ingredients) at the cost of dis-including predictor variables with small but real effects. An investigation interested in something closer to pure prediction might choose an index more tolerant of low- or no-effect covariates. Here, our use of ΔSIC has the advantage of allowing the structure of evidential analysis to be portrayed, studied, and executed in the context of standard linear model theory. Use of other evidence functions will entail a greater reliance on computer simulation [8].

3. Hypothesis tests and evidential analysis in normal linear models

Many standard analyses in applied statistics are contained within the family of normal linear models with fixed effects. The normal linear fixed effects model takes data vector $\boldsymbol{y}$ $(n \times 1)$ to have arisen from a multivariate normal distribution with mean vector $\boldsymbol{X\beta}$ and variance-covariance matrix $\sigma^2\boldsymbol{I}$, where $n$ is the number of observations, $\boldsymbol{X}$ is a full-column-rank design matrix $(n \times r)$, $\boldsymbol{\beta}$ is a vector $(r \times 1)$ of parameters, $\boldsymbol{I}$ is the identity matrix $(n \times n)$, and $\sigma^2$ is a positive scalar parameter. The individual observations in the data vector $\boldsymbol{y}$ under the normal linear model are independent but generally not identically distributed, having different means as prescribed by the design matrix $\boldsymbol{X}$. The likelihood function for the parameters given the observed data $\boldsymbol{y}$ is a multivariate normal pdf evaluated at $\boldsymbol{y}$:

$$L(\boldsymbol{\beta}, \sigma^2) = (2\pi\sigma^2)^{-n/2}\exp\left[-\frac{(\boldsymbol{y}-\boldsymbol{X\beta})'(\boldsymbol{y}-\boldsymbol{X\beta})}{2\sigma^2}\right]. \quad (11)$$

The estimation and NP hypothesis testing material quoted here come from standard results in the theory of linear models (for example, [10, 11]). A basic well-known result gives the maximum likelihood (ML) estimates of $\boldsymbol{\beta}$ and $\sigma^2$ as

$$\widehat{\boldsymbol{\beta}} = (\boldsymbol{X}'\boldsymbol{X})^{-1}\boldsymbol{X}'\boldsymbol{y}\,, \quad (12)$$

$$\hat{\sigma}^2 = \left(\boldsymbol{y} - \boldsymbol{X}\widehat{\boldsymbol{\beta}}\right)'\left(\boldsymbol{y} - \boldsymbol{X}\widehat{\boldsymbol{\beta}}\right)/n\,. \quad (13)$$

For many inferential purposes, the unbiased estimate of $\sigma^2$ given by

$$\tilde{\sigma}^2 = n\hat{\sigma}^2/(n-r) \quad (14)$$

is preferred, as the ML estimate of $\sigma^2$ can substantially underestimate uncertainty when the number of observations is not sufficiently greater than the number of estimated parameters.

For the normal linear fixed effects model, the generalized likelihood ratio statistic (Eq. 6) for testing a constrained null vs unconstrained alternative hypothesis is a monotone function of an F statistic with a noncentral F distribution. An evidence function for such a model comparison based on ΔSIC then becomes a monotone function of that F statistic. Thus, the noncentral F distribution will be the go-to distribution for approaching a linear model problem as an evidential analysis. The noncentral F is a heavy-tailed distribution on the positive real line (Figure 1).

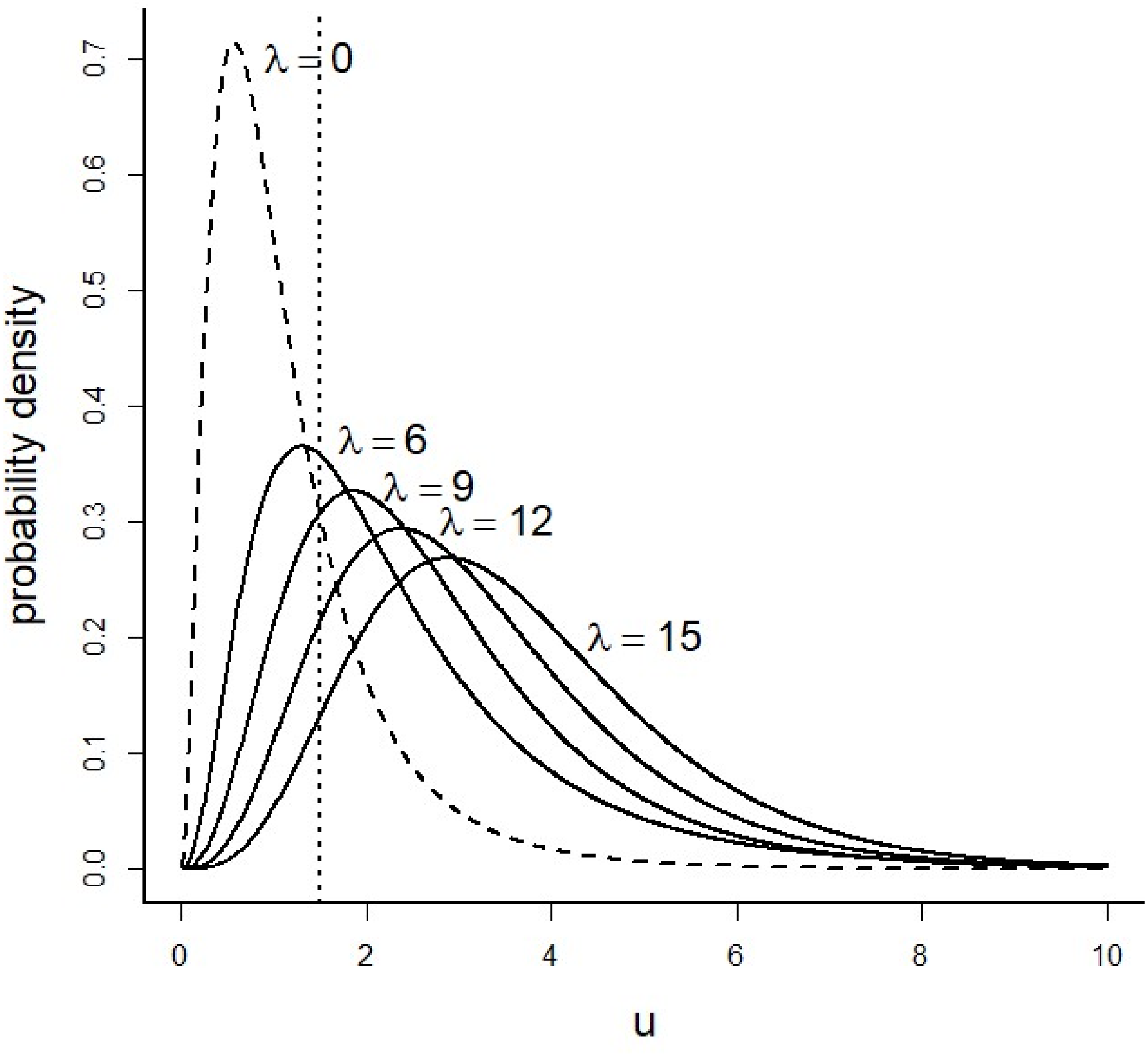

Figure 1. Probability density functions (solid curves) of the noncentral F$(q, n - r, \lambda)$ distribution for various values of sample size $n$ and the noncentrality parameter $\lambda$, as represented in the formula for $f(u)$ in the text, Eq. 30. Here $\lambda = n\delta^2$, is in the common form of a simple experimental design, in which $n$ is the number of observations and $\delta^2$ is a generalized squared per-observation effect size. The cumulative distribution function of the noncentral F distribution, exemplified here as the area under each density curve to the left of the dashed vertical line, is a monotone decreasing function of $n$. Here $q = 6$, $r = 12$, $\delta^2 = .25$, and $n$ has the values $24$, $36$, $48$, and $60$. Dashed curve is the density function for the $F(q, n - r, \lambda)$ distribution with $n = 24$ and $\delta^2 = 0$ (central F distribution).

Specifically, if $\hat{L}_1$ and $\hat{L}_2$ are the maximized likelihoods under the null and alternative models respectively, then

$$G^2 = -2\log\left(\frac{\hat{L}_1}{\hat{L}_2}\right) = n\log\left(1 + \frac{q}{n-r}F\right), \qquad (15)$$

where $F$ is the F statistic, and $r - q$ is the number of unknown parameters in the mean of the null model (i.e., $q$ linear constraints are being imposed on the parameters in $\boldsymbol{\beta}$ in the null model, so that $q$ is the difference of the number of unknown parameters in the alternative and null models). If $q = 1$, then $F = T^2$, where $T$ has a noncentral T distribution.

From the relationship (Eq. 7) between ΔSIC and $G^2$, the evidence function for normal linear models based on ΔSIC becomes

$$\Delta\text{SIC} = G^2 - q\log(n) = n\log\left(1 + \frac{q}{n-r}F\right) - q\log(n)\,. \qquad (16)$$

This evidence function based on ΔSIC is easily calculated from the information provided by analysis of variance tables in standard statistical software or from straightforward commands in computer programming languages.

4. Neyman Pearson hypothesis test formulations

Two different formulations of NP hypothesis tests in linear models are convenient for alternative study with evidential analysis. The first formulation (A) makes it easy to ask which parameters are not 0. The second formulation (B) makes it easy to identify differences between parameters.

(A) This formulation constructs a null hypothesis in which one or more of the parameters in the vector $\boldsymbol{\beta}$ are set to zero, such as dropping one or more variables in a multiple regression. Write $\boldsymbol{X}$ and $\boldsymbol{\beta}$ as partitioned matrixes in the form

$$\boldsymbol{\beta} = \begin{bmatrix} \boldsymbol{\beta}_1 \\ \boldsymbol{\beta}_2 \end{bmatrix} \text{, and } \boldsymbol{X} = (\boldsymbol{X}_1, \mathbf{X}_2) , \tag{17}$$

so that

$$\boldsymbol{X\beta} = \boldsymbol{X}_1\boldsymbol{\beta}_1 + \boldsymbol{X}_2\boldsymbol{\beta}_2 . \tag{18}$$

Here the vector $\boldsymbol{\beta}_2$ $(q \times 1)$ contains the parameters to be set to zero under the null model. Also, the matrix $\boldsymbol{X}_2$ $(n \times q)$ contains the columns of $\boldsymbol{X}$ corresponding to the $\boldsymbol{\beta}_2$ parameters that are to be dropped under the null model, with $\boldsymbol{\beta}_1$ $((r-q) \times 1)$ and $\boldsymbol{X}_1$ $(n \times (r-q))$ carrying the model components to be retained under the null model. The two models (hypotheses $\mathrm{H}_1$ and $\mathrm{H}_2$, which we often refer to as model 1 and model 2) are

$$\mathrm{H}_1\colon\ \boldsymbol{\beta}_2 = \mathbf{0} , \tag{19}$$

$$\mathrm{H}_2\colon\ \boldsymbol{\beta}_2 \neq \mathbf{0} . \tag{20}$$

From Eqs. 12 and 13 above we have the ML (least squares) estimate of $\boldsymbol{\beta}$ and $\sigma^2$ under the unrestricted alternative model $\mathrm{H}_2$. Model $\mathrm{H}_1$ similarly provides its own ML estimates as

$$\widehat{\boldsymbol{\beta}}_1^* = \left(\boldsymbol{X}_1{}'\boldsymbol{X}_1\right)^{-1}\boldsymbol{X}_1{}'\boldsymbol{y} ; \tag{21}$$

$$\hat{\sigma}_1{}^2 = \left(\boldsymbol{y} - \boldsymbol{X}_1\widehat{\boldsymbol{\beta}}_1^*\right)' \left(\boldsymbol{y} - \boldsymbol{X}_1\widehat{\boldsymbol{\beta}}_1^*\right)/n . \tag{22}$$

The generalized likelihood ratio statistic for testing $\mathrm{H}_1$ versus $\mathrm{H}_2$, via Eq. 15, reduces in this formulation to a monotone function of an F statistic of the form

$$F = \frac{\left(\widehat{\boldsymbol{\beta}}'\boldsymbol{X}'\boldsymbol{y} - \widehat{\boldsymbol{\beta}}_1^{*\prime}\boldsymbol{X}_1{}'\boldsymbol{y}\right)/q}{(\boldsymbol{y}'\boldsymbol{y} - \widehat{\boldsymbol{\beta}}'\boldsymbol{X}'\boldsymbol{y})/(n-r)} . \tag{23}$$

The F statistic (pre-data) has a noncentral F distribution, written $F \sim \mathrm{F}(q, n-r, \lambda)$, with numerator and denominator degrees of freedom given respectively by $q$ and $n-r$ and noncentrality parameter $\lambda$ given by

$$\lambda = \frac{\boldsymbol{\beta}_2{}'\left(\boldsymbol{X}_2{}'\boldsymbol{X}_2 - \boldsymbol{X}_2{}'\boldsymbol{X}_1(\boldsymbol{X}_1{}'\boldsymbol{X}_1)^{-1}\boldsymbol{X}_1{}'\boldsymbol{X}_2\right)\boldsymbol{\beta}_2}{\sigma^2} . \tag{24}$$

Under the null model ($H_1$), $\lambda = 0$, and an ordinary central F statistic applies.

(B) The second formulation of hypothesis testing is convenient for testing one or more linear contrasts among the parameters in $\boldsymbol{\beta}$. The two models are given by

$$H_1\colon\ \boldsymbol{L\beta} = \boldsymbol{h}\,, \tag{25}$$

$$H_2\colon\ \boldsymbol{L\beta} \neq \boldsymbol{h}\,. \tag{26}$$

Here $\boldsymbol{L}$ is a $q \times r$ matrix of known constants, and $\boldsymbol{h}$ is a $q \times 1$ vector of known constants (frequently zeros). The F statistic for testing $H_1$ versus $H_2$ becomes

$$F = \frac{(\boldsymbol{L}\hat{\boldsymbol{\beta}}-\boldsymbol{h})'\left(\boldsymbol{L}(\boldsymbol{X}'\boldsymbol{X})^{-1}\boldsymbol{L}'\right)^{-1}(\boldsymbol{L}\hat{\boldsymbol{\beta}}-\boldsymbol{h})/q}{(\boldsymbol{y}'\boldsymbol{y}-\hat{\boldsymbol{\beta}}'\boldsymbol{X}'\boldsymbol{y})/(n-r)}\,, \tag{27}$$

with $F \sim \mathrm{F}(q, n-r, \lambda)$, where the noncentrality parameter is now

$$\lambda = \frac{(\boldsymbol{L\beta}-\boldsymbol{h})'\left(\boldsymbol{L}(\boldsymbol{X}'\boldsymbol{X})^{-1}\boldsymbol{L}'\right)^{-1}(\boldsymbol{L\beta}-\boldsymbol{h})}{\sigma^2}\,. \tag{28}$$

Both versions (Eqs. 23 and 27) of the $F$ statistic are algebraically equivalent to the familiar "reduction in variance" form given by

$$F = \left(\frac{n-r}{q}\right)\left(\frac{\hat{\sigma}_1{}^2-\hat{\sigma}^2}{\hat{\sigma}^2}\right). \tag{29}$$

5. Closer look at noncentrality

One role of the noncentral F distribution in an evidential analysis is to help in selecting the evidence cutoff values $k_1$ and $k_2$. For pre-data design purposes the threshold values $k_1$ and $k_2$ can be set under the assumptions that the governing vector parameter value lies in the null model ($H_1$) or in the alternative model ($H_2$), respectively. In evidential practice, "in the null model" is taken to mean "within some ignorable distance of the null model", that is, small but negligible departures of $\boldsymbol{\beta}_2$ from 0 are allowed as an adequate specification of the data generating mechanism under $H_1$. Here evidential practice departs substantially from NP hypothesis testing. While NP testing sets the Type 1 error probability $\alpha$ based on the literal parameter constraint represented by $\mathrm{H}_1$, evidential analysis, in setting the probability of misleading evidence $M_1$, uses the practical meaning of $\mathrm{H}_1$ as indicating model components that can be ignored for the purposes at hand. Thus, the correct distribution for comparing the two models is a noncentral F distribution instead of a central F as in NP hypothesis testing. The two probabilities of misleading evidence will be related to two tail areas under an appropriate noncentral F distribution. The probability $M_1$ (Eq. 8) is the area to the right of $k_2$, and $M_2$ (Eq. 9) is the area to the left of $k_1$, under the distribution of ΔSIC, which in turn is related to an F($q$, $n - r$, $\lambda$) distribution via Eq. 16. To set $k_1$ and $k_2$, the appropriate value of $\lambda$ in the F($q$, $n - r$, $\lambda$) distribution will depend on the investigator's decision about the zone of indifference for $\mathrm{H}_1$: a zone of parameter values representing negligible departures from $\mathrm{H}_1$ for purposes of model selection. In this sense, the process of setting $k$ values resembles power calculations in NP testing: in NP testing one picks a sample size and a study design, under a given "effect size" desired to be detected by the study and under a given Type 1 error rate, so as to make the Type 2 error rate (invariably denoted $\beta$, not to be confused with the usual notation for the parameter vector in the mean of a linear model) as small as desired. The main difference in evidential analysis is that design, sample size, and values of $k_1$ and $k_2$ are picked so as to make both misleading error rates ($M_1$ and $M_2$) as small as desired. As well, some studies might focus on the probabilities $W_1$ and $W_2$ of weak evidence, which will be areas under the distribution of ΔSIC between $k_1$ and $k_2$.

To avail themselves of the advantages of evidential analysis over standard approaches in the statistical canon, data analysts will need to become more familiar with noncentrality. The noncentral versions of the F, t, and chisquare distributions that figure in statistical hypothesis testing for power calculations and experimental design are used in evidential analysis in pre- and post-data roles. Pre-data, with the noncentral distributions one can set the evidence thresholds ($k$ values), misleading evidence probabilities ($M$ values), or sample sizes necessary to attain whatever $k$ and $M$ values are sought. Post-data, the noncentral distributions provide local calculations for just how secure (or insecure) the obtained evidence is, via assessment of uncertainties in the analysis.

The task of making noncentral distributions part of statistical routine is not straightforward. Users of applied statistics have mostly interacted with noncentral distributions through the complex multidimensional graphs in the back of experimental design textbooks. Students of linear model theory may learn the noncentral distributions in computationally nonfriendly ways, such as expressing the noncentrality parameter in terms of the projection of one space on another. Exacerbating the project is that different books parameterize the noncentral distributions in different ways. While excellent software is available for calculations with the noncentral distributions, the accompanying documentation can be opaque about the exact details of how the noncentrality parameter is defined.

We attempt a standardization here, at least for clarifying how calculations are done in the example we present. The noncentral F distribution used in the above formulas, abbreviated by F($\nu_1, \nu_2, \lambda$), has pdf given by

$$p(u) \;=\; \sum_{j=0}^{\infty} \frac{e^{-\frac{\lambda}{2}}\left(\frac{\lambda}{2}\right)^{j}}{j!} \frac{\Gamma\left(\frac{\nu_1}{2}+j+\frac{\nu_2}{2}\right)}{\Gamma\left(\frac{\nu_1}{2}+j\right)\Gamma\left(\frac{\nu_2}{2}\right)} \left(\frac{\nu_1}{\nu_2}\right)^{\left(\frac{\nu_1}{2}+j\right)} \left(\frac{\nu_2}{\nu_2+\nu_1 u}\right)^{\left(\frac{\nu_1}{2}+j+\frac{\nu_2}{2}\right)} u^{\left(\frac{\nu_1}{2}+j-1\right)}\,, \qquad (30)$$

where $u$ is a positive real variate value for a random variable with a noncentral F random distribution, $\nu_1$ and $\nu_2$ are positive integers, and $\lambda$ is a nonnegative real quantity termed the "noncentrality parameter." The formula is a weighted sum (mixture) of a countably infinite number of central F distributions, with the weights being Poisson probabilities from a Poisson distribution with a mean of $\lambda/2$. The curious Poisson terms, as discrete probabilities, seem out of place in the pdf formula, because no explicit Poisson process is evident in data gathering, but the terms arise fundamentally from the tail-probability relationship between the gamma (chisquare) and the Poisson distributions [20].

A main point of confusion occurs because some texts and software products define the noncentrality parameter to be $\lambda/2$ instead of $\lambda$. An easy way to check which definition is used in a computer program is to simulate many(!) values from its noncentral F distribution and calculate the sample mean to compare with the distribution mean, noting that the expected value for the distribution with pdf above is

$$E(F) = \frac{\nu_2(\nu_1+\lambda)}{\nu_1(\nu_2-2)}, \tag{31}$$

provided $\nu_2 > 2$. For instance, with $\nu_1 = 1$, $\nu_2 = 3$, and $\lambda = 2$, the expected value for the above distribution is 9. If, however, the noncentrality parameter is defined in the program as $\xi = \lambda/2$, then using $\nu_1 = 1$, $\nu_2 = 3$, and $\xi = 2$ in the programmed distribution will produce random variates with expected values of 15. One can expect substantial variability in the sample mean, so a large sample of variates, say 10,000, should be generated. We note that the variance of the noncentral F distribution does not exist unless $\nu_2 > 4$, but the law of large numbers (convergence of the sample mean to the distribution mean) does not require the existence of the variance. The F distribution functions from the stats package in R (as of version 4.4.0) use a noncentrality parameter, ncp, defined as we have.

Four aspects of the noncentrality parameter are noteworthy for evidential analysis: 1) Sample size $n$ does not appear explicitly in the general formula(s) for $\lambda$ (Eqs. 24, 28) but is rather wrapped implicitly into the study design and hypotheses in question. In some cases, $n$ will explicitly pop out algebraically, in other cases, it will not. 2) The noncentrality quantity $\lambda$ is often represented as a product of $n$ and a per-observation "effect size" (relative to $\sigma^2$). The effect size has $\sigma^2$ in the denominator and a function of those $\boldsymbol{\beta}$ parameters constrained under model 1 in the numerator. The numerator of the effect size measures the departure of the true parameters from their constrained values under model 1 and is the squared difference of the parameter from its constrained value if model 1 posits a constraint on a single scalar parameter. However, if the model 1 constraint is on two or more parameters, the effect size numerator will be a quadratic form of the vector of parameters in question, a type of generalized squared departure distance, with the quadratic form matrix being a complicated amalgamation of the study design characteristics. The effect size aspect of $\lambda$ is made more explicit below in Section 7 below. 3) The complexity of $\lambda$ builds up rapidly as the number of columns of $\boldsymbol{X}$ increase. In many cases symbolic simplification is not possible, and numerical calculation will be necessary. Algebraic variants of the formulas for $\lambda$ are numerous, and there might be cases in which computer symbolic algebra yields insights. 4) The classical effect size as represented by $\lambda$ is related to, but in general not equal to, the KL divergence between model 1 and model 2 (see Section 6 below). Aspects of the study design enter into $\lambda$ along with KL divergence, because $\lambda$ embodies the propensity for errors, that is, the ability of the number of observations as apportioned in the design to inform about the KL divergence. Poor design decreases the effectiveness of data in detecting a given KL divergence.

## 6. Relationship of noncentrality to Kullback-Leibler divergence

The Kullback-Leibler (KL) divergence measure is a key underpinning of ordinary evidential analysis, although alternative evidential systems can be constructed upon other distribution divergence measures. Evidential analysis seeks to determine which of two or more models is a better description of the probabilistic mechanism generating the data, and the KL divergences of one model from another, and of both models from the actual data generating mechanism, are central targets of inference [8].

Suppose $f_1(\boldsymbol{y})$ is the pdf of an $n$-dimensional multivariate normal distribution with a mean vector given by $\boldsymbol{\mu}_1$ and variance-covariance matrix given by $\boldsymbol{\Sigma}_1$, and suppose $f_2(\boldsymbol{y})$ is also an $n$-dimensional multivariate normal pdf with mean vector $\boldsymbol{\mu}_2$ and variance-covariance matrix $\boldsymbol{\Sigma}_2$. The KL divergence of $f_2$ from $f_1$ is the expected value of the log-likelihood ratio, $\log[f_1(\boldsymbol{Y})/f_2(\boldsymbol{Y})]$ , with the expectation taken with respect to $f_1$. For two (nonsingular) multivariate normal pdfs, the following standard result is listed in many references:

$$K(f_1\,, f_2) = \frac{1}{2}\left[\mathrm{tr}\left(\boldsymbol{\Sigma}_1{}^{-1}\boldsymbol{\Sigma}_2\right) - n + (\boldsymbol{\mu}_1 - \boldsymbol{\mu}_2)'\boldsymbol{\Sigma}_1{}^{-1}(\boldsymbol{\mu}_1 - \boldsymbol{\mu}_2) + \log\left(\frac{|\boldsymbol{\Sigma}_1|}{|\boldsymbol{\Sigma}_2|}\right)\right]. \qquad (32)$$

The KL divergence of $f_1$ from $f_2$ simply reverses the subscripts.

Under the first formulation of hypothesis testing for linear models (Eqs. 19, 20), $\boldsymbol{\Sigma}_1 = \boldsymbol{\Sigma}_2 = \sigma^2\boldsymbol{I}$, $\boldsymbol{\mu}_1 = \boldsymbol{X}_1\boldsymbol{\beta}_1$, and $\boldsymbol{\mu}_2 = \boldsymbol{X}_1\boldsymbol{\beta}_1 + \boldsymbol{X}_2\boldsymbol{\beta}_2$. Substituting, we find that

$$K(f_1\,, f_2) = \frac{\boldsymbol{\beta}_2{}'\boldsymbol{X}_2{}'\boldsymbol{X}_2\boldsymbol{\beta}_2}{2\sigma^2}. \qquad (33)$$

KL divergences are not generally symmetric, but interestingly in this normal distribution case, the KL divergence of $f_1$ from $f_2$ is then identical to $K(f_1\,, f_2)$. The noncentrality parameter (Eq. 24) written in terms of $K(f_1\,, f_2)$ becomes

$$\lambda = 2K(f_1\,, f_2)\; - \;\frac{\boldsymbol{\beta}_2{}'\boldsymbol{X}_2{}'\boldsymbol{X}_1(\boldsymbol{X}_1{}'\boldsymbol{X}_1)^{-1}\boldsymbol{X}_1{}'\boldsymbol{X}_2\boldsymbol{\beta}_2}{\sigma^2}. \qquad (34)$$

Here $2K(f_1\,, f_2) = K(f_1\,, f_2) + K(f_2\,, f_1)$ is the symmetric KL distance (a true distance measure).

## 7. Analysis and study design

While the broad target of evidential analysis is difference of departures of $f_1$ and $f_2$ from some unknown true model $g$ [4, 8], the design of the study can be based upon the assumption that one of the model candidates adequately approximates $g$, provided the selected model is subsequently subjected to model quality probes. Under the adequate model assumption, misleading evidence probabilities will depend on the noncentral F distribution with noncentrality parameter $\lambda$ through the relationship between evidence function $\Delta\text{SIC}$ and test statistic $F$ (Eq. 16). The distribution of $\Delta\text{SIC}$ has the properties that the misleading evidence probabilities $M_2$ and $M_1$ both go toward zero as sample size increases, once the values of $k_1$ and $k_2$ are set. However, the planned values of $k_1$, $k_2$, $M_2$, and $M_1$ will depend on how observations are allocated in the study design. In general, as the planned value of $n$ increases, $\lambda$ will increase, and the distance between the planned values for $k_1$ and $k_2$ will decrease. The term subtracted from $2K$ on the right side of Eq. 34 represents a diminishment of $\lambda$ due to study design characteristics, a "design load" causing increase of the misleading error probabilities $M_2$ and $M_1$ for fixed values of $k_1$ and $k_2$.

One strategy for setting evidential benchmarks in simple designs starts by writing $\lambda$ as a product of $n$ and a per-observation relative effect size. In expressions such as Eq. 24, $\lambda$ is a ratio, in which the numerator is a scalar measure of information of interest to the model selection in the form of a generalized "squared" departure of model 2 parameters from model 1 parameters, taking into account the intrinsic limitations in the study design. The denominator of $\lambda$ is the per-observation variance representing the general noise level clouding the inference(s) in question. Thus, we see that $\lambda$ as represented by Eq. 24 is in the following form:

$$\lambda = \frac{\boldsymbol{\beta}_2{}'\boldsymbol{A}\boldsymbol{\beta}_2}{\sigma^2}\,. \qquad (35)$$

Here $\boldsymbol{A}$ is the known matrix in the numerator of Eq. 24 producing the quadratic form in $\boldsymbol{\beta}_2$. In many simpler designs, such as one- or two- way analysis of variance, $\left(\frac{1}{n}\right)\boldsymbol{\beta}_2{}'\boldsymbol{A}\boldsymbol{\beta}_2$ is constant or approaches a constant as $n$ becomes large, containing for instance the proportions of observations allocated by design to different treatment combinations. If the design is complex, requiring observations allocated among many treatment combinations, the convergence to a constant might require a large $n$ value, and using the form of $\lambda$ with the full expression for design load (Eq. 24) is recommended. For planning simpler designs, one can write $\lambda$ as the product of $n$ and a ratio:

$$\lambda = n\frac{\left(\frac{1}{n}\right)\boldsymbol{\beta}_2{}'\boldsymbol{A}\boldsymbol{\beta}_2}{\sigma^2} = n\delta^2 \ . \qquad (36)$$

The ratio numerator becomes a per-observation, generalized "squared" departure of model 2 from model 1. The denominator is the per-observation variance. The ratio itself is the relative effect size (often termed just the “effect size”). One can predesignate a value of $\left(\frac{1}{n}\right)\boldsymbol{\beta}_2{}'\boldsymbol{A}\boldsymbol{\beta}_2$ equal to some multiple of $\sigma^2$, say $\left(\frac{1}{n}\right)\boldsymbol{\beta}_2{}'\boldsymbol{A}\boldsymbol{\beta}_2 = \delta^2\sigma^2$, so that $\lambda = n\delta^2$. Here $\delta\sigma$ represents the maximum allowable size of $\sqrt{\boldsymbol{\beta}_2{}'\boldsymbol{A}\boldsymbol{\beta}_2/n}$ (per observation departure of model 2 from model 1) considered consistent with model 1. Thus, for designing a study for evidential analysis, one can fix the numerator of the relative effect size in terms of a smallest multiple of $\sigma$ that one desires to detect with given probabilities of misleading evidence; for instance $\lambda = n(.5\sigma)^2/\sigma^2 = .25n$ would represent half of a standard deviation as the largest tolerable departure of parameters consistent with model 1. In other words, if one has a preplanned sample size $n$ and one wanted to wanted to pick values of $k_1$ and $k_2$ values, or if one wanted to plan the sample size with predesignated $k_1$ and $k_2$ values, then to detect a per-observation departure of model 2 from model 1 that was 50 percent of the observation standard deviation $\sigma$, with misleading evidence probabilities $M_1$ and $M_2$ no larger than desired, then one would work with the tail probabilities of the noncentral F distribution with $\lambda = .25n$. The calculations flow from the fact that the study outcome will be deemed strong evidence for $\mathrm{H}_1$ if

$$\Delta\mathrm{SIC} = n \log\left[1 + \left(\frac{q}{n-r}\right) F\right] - q \log(n) < k_1 , \tag{37}$$

strong evidence for $\mathrm{H}_2$ if

$$\Delta\mathrm{SIC} = n \log\left[1 + \left(\frac{q}{n-r}\right) F\right] - q \log(n) > k_2 , \tag{38}$$

and weak or inconclusive evidence otherwise.

For example, suppose sample size $n$ has been predesignated, and one wants to set a value of $k_1$ that will render the probability $M_2$ of misleading evidence for model $f_1$ (i.e. $\mathrm{H}_1$) to be no larger than some specified constant, say $\gamma_2$; perhaps $\gamma_2 = .05$. Suppose the smallest multiple of $\sigma$ one designates to delineate between models $f_1$ and $f_2$ is $\delta$; perhaps $\delta = .5$. Under model $f_2$, the probability of misleading evidence for $f_1$ is

$$M_2 = P(\Delta\mathrm{SIC} < k_1 \mid f_2) = P\left(n \log\left[1 + \left(\frac{q}{n-r}\right) F\right] - q \log(n) < k_1 \mid f_2\right)$$

$$= P\left(F < \frac{n-r}{q}\left[n^{q/n} e^{k_1/n} - 1\right] \mid f_2\right) . \tag{39}$$

One sets $k_1$ to ensure that $M_2 \leq \gamma_2$, that is, to ensure that the probability of misleading evidence is less than some fixed known value. Find the $\gamma_2$ quantile, say $\psi_1$, of a noncentral $\mathrm{F}(q, n-r, n\delta^2)$ distribution. That noncentral $\mathrm{F}(q, n-r, n\delta^2)$ distribution is on the boundary between the two models and will produce the maximum value of $M_2$ for a given $n$ (the left tail of the $\mathrm{F}(q, n-r, n\delta^2)$ distribution decreases as $\delta^2$ increases). The quantile function for the noncentral F distribution is available as the qf( , , ) function in R. The value of $k_1$ is obtained by equating $\psi_1$ and $\left(\frac{n-r}{q}\right)\left[n^{q/n} e^{k_1/n} - 1\right]$, producing

$$k_1 = n \log\left[1 + \left(\frac{q}{n-r}\right)\psi_1\right] - q\log(n)\ . \qquad (40)$$

The above value of $k_1$ then guarantees that $M_2 \le \gamma_2$ .

The value of $k_2$ is set through similar reasoning. Under model $f_1$, the probability of misleading evidence for $f_2$ is

$$M_1 = P(\Delta\text{SIC} > k_2 \mid f_1) = P\left(n\log\left[1 + \left(\frac{q}{n-r}\right)F\right] - q\log(n) > k_2 \mid f_1\right)$$

$$= P\left(F > \left(\frac{n-r}{q}\right)\left[n^{q/n}e^{k_2/n} - 1\right] \mid f_1\right)\ . \qquad (41)$$

The value of $k_2$ is picked to ensure that $M_1 \le \gamma_1$ (say, $\gamma_1 = 0.05$). Let $\psi_2$ be the $(1-\gamma_1)$ quantile of the noncentral F$(q, n-r, n\delta^2)$ distribution, the boundary between the two models again providing the largest misleading error probability. Take

$$k_2 = n\log\left[1 + \left(\frac{q}{n-r}\right)\psi_2\right] - q\log(n) \qquad (42)$$

to guarantee that $M_1 \le \gamma_1$.

If instead $k_1$ is predesignated (predesignated evidence threshold for strong evidence for model 1), and one wants $M_2$ say to be less than or equal to $\gamma_2$, one would find the value of $n$ that makes the left tail of an F$(q, n-r, n\delta^2)$ distribution below $\psi_1$ equal to $\gamma_2$, where

$$\psi_1 = \left(\frac{n-r}{q}\right)\left[n^{q/n}e^{k_1/n} - 1\right]\ . \qquad (43)$$

Obtaining the value of $n$ would be a numerical root-finding calculation, using a software routine for the cdf of a noncentral F distribution such as pf( , , ) in R. An approximate $n$ value can be obtained by graphing the noncentral F cdf versus a range of $n$ values.

Likewise for predesignated $k_2$: to attain $M_1 \le \gamma_1$, find the value of $n$ that makes the right tail of an F$(q, n-r, n\delta^2)$ distribution above $\psi_2$ equal to $\gamma_1$, where

$$\psi_2 = \left(\frac{n-r}{q}\right)\left[n^{q/n}e^{k_2/n} - 1\right]\ . \qquad (44)$$

Interestingly, if $\gamma_1$ and $\gamma_2$ are taken to be equal, say at a value of $\gamma$, one can validly claim in a frequentist sense (under the assumption of correct model specification) that "the probability of misleading evidence for this study is no larger than $\gamma$", provided the model family represented by model 2 is adequate. The same can be stated for whichever of $\gamma_1$ and $\gamma_2$ is the largest.

8. Post-data evaluations

Once the observations are recorded and the evidence function has made its trichotomous choice between $f_1$, $f_2$, or neither, some post-data analyses are informative. Suppose Δsic is the realized value of ΔSIC, the lower case denoting an outcome, not a random variable. Suppose the outcome wound up with evidence, strong or weak, favoring one of the models. One can calculate an analog of a $p$-value in hypothesis testing by determining how extreme is the value of Δsic under the disfavored model. One poses the question: if the experiment were repeated with the disfavored model $f_i$ generating the data, what is the largest probability that the evidence would be as misleading as Δsic? The probability in question, denoted $P_i$, is obtained from the noncentral F$(q, n-r, \lambda)$ distribution, with the noncentrality parameter $\lambda$ set to $n\delta^2$. The value $\delta$ is the border of the parameter space tolerance region separating the two models and will yield the largest probabilities of misleading evidence. Note that if the true value of the effect size is near $\delta$, the distribution of ΔSIC will be centered around 0; the left tail will decrease when the true effect size shifts toward model 2 (ΔSIC distribution centered on the positive line), and the right tail will decrease when the true effect size shifts in favor of model 1 (ΔSIC distribution centered on the negative line). Let $\Psi(f, q, n-r, n\delta^2)$ denote the cdf of a noncentral F$(q, n-r, n\delta^2)$ distribution. If model 1 is favored by Δsic, we have

$$P_2 = \max \mathrm{P}(\Delta\mathrm{SIC} \le \Delta\mathrm{sic} \mid f_2) = P\left(n \log\left[1 + \left(\frac{q}{n-r}\right) F\right] - q \log(n) \le \Delta\mathrm{sic}\right)$$

$$= P\left(F \leq \frac{n-r}{q}\left[n^{q/n}e^{\Delta\mathrm{sic}/n} - 1\right]\right) = \Psi\left(\frac{n-r}{q}\left[n^{q/n}e^{\Delta\mathrm{sic}/n} - 1\right], q, n-r, n\delta^2\right). \qquad (45)$$

If model 2 is favored by Δsic, we have

$$P_1 = \max \mathrm{P}(\Delta\mathrm{SIC} \geq \Delta\mathrm{sic} \mid f_1) = P\left(n \log\left[1 + \left(\frac{q}{n-r}\right)F\right] - q\log(n) \geq \Delta\mathrm{sic}\right)$$

$$= P\left(F \geq \frac{n-r}{q}\left[n^{q/n}e^{\Delta\mathrm{sic}/n} - 1\right]\right) = 1 - \Psi\left(\frac{n-r}{q}\left[n^{q/n}e^{\Delta\mathrm{sic}/n} - 1\right], q, n-r, n\delta^2\right)$$

$$= 1 - P_2\,. \qquad (46)$$

The value of Δsic can be used for a post data determination of the smallest value of $\delta$ for which there is strong evidence under model 1, or the largest value of $\delta$ under which there is strong evidence for model 2. In other words, the NP dilemma of what constitutes evidence for the null hypothesis can be disentangled and studied. One takes the expressions for $P_2$ and $P_1$ above (Eqs. 45, 46) and calculates them as functions of $\delta$. The levels of $\delta$ that correspond to "strong" levels of $P_2$ or $P_1$ (however small as designated by the investigator) are the per-observation effect sizes warranted by the data.

The prevailing distribution of ΔSIC can be estimated. A straightforward method of estimating the distribution of ΔSIC is with bootstrapping, either parametric or some form of nonparametic bootstrapping. First, we describe a parametric bootstrap. We subsequently delve into an approach to nonparametric bootstrapping.

The distribution from which the observations arose can be estimated as a multivariate normal distribution with a mean vector of $\boldsymbol{X}\widehat{\boldsymbol{\beta}}$ and variance-covariance matrix of $\tilde{\sigma}^2\boldsymbol{I}$ (Eqs. 12 and 14; the ML-unbiased estimate of $\sigma^2$ is much preferred here to the ML estimate). Of course, the observations under the modeling framework are independent, and the joint distribution can be alternatively estimated as a product-normal with the above parameter estimates; the multivariate form is merely a convenience toward coding brevity for simulation in computer languages that have matrix calculations and multivariate distributions built in.

The idea is to generate $n_{\mathrm{B}}$ bootstrap data sets (perhaps 1000 or more) from the estimated joint distribution. Each bootstrap data set consists of a $n \times 1$ vector. From each bootstrap data set, one refits the 2 models and calculates a value of ΔSIC, along with any other statistics of interest, such as the estimated KL divergence (Eq. 33 using $\widehat{\boldsymbol{\beta}}$ and $\tilde{\sigma}^2$ values calculated from each bootstrap sample). A graph of the empirical distribution function (EDF) of the bootstrap ΔSIC values (or other statistics of interest), along with calculated benchmarks such as percentiles, provide insight into the uncertainty accompanying the analyses and conclusions about the model comparison. The EDF can be smoothed with a kernel distribution estimator of the pdf if desired. Such smoothing allows better estimation of probabilities (areas under pdf curves). Univariate local polynomial kernel density estimators [21] are effective, particularly with data that is bounded (like nonnegative distributions), or has long tails, or both.

Nonparametric bootstrapping of linear models requires attention to the model structure. Identical distribution (as opposed to independence) is only found within factor combination cells. In regression-style studies with quantitative predictor variables, there is often only one observation for each combination of predictor variable levels. For inference on regression, one can use a form of semiparametric bootstrapping in which the residuals from the fitted model are sampled with replacement (instead of sampling from the estimated normal distribution model as in parametric bootstrapping) and added to the model-estimated expected values to construct each bootstrap data set.

In ANOVA-style studies with categorical predictor variables there are ideally multiple observations within each combination of treatment levels. Multiple observations per cell allows some exploration of uncertainty of evidential analysis in a fashion closer to a model-free ideal. There are various approaches to nonparametric bootstrapping of ANOVA data, usually involving stratification of the resampling process within cells. We presently recommend one method, described in what follows. For studies with mixed quantitative and categorical predictor variables the semiparametric approach can be used, or possibly hybrid approaches can be devised.

The bootstrap we recommend for ANOVA-style studies has been shown to be reliable and robust [22, 23] and involves bootstrapping variance-inflated, median-centered residuals within cells (treatment level combinations). The bootstrap is accomplished with the following steps: 1) Calculate the median response, $\widetilde{M}_i$ , for each cell $i$. 2) Calculate residuals in each cell $i$ as $\left(y_{i,j} - \widetilde{M}_i\right)s_i$, where $s_i = \sqrt{[n_i/(n_i - 1)]}$ is a scaling factor for the residuals of cell $i$, with $n_i$ being the number of observations in the cell. The scaling factor inflates the expected sample variance of the residuals in each cell to equal the population variance for that cell. 3) Create a bootstrapped response vector $\boldsymbol{y}^*$ by adding to each cell median a bootstrapped set of variance-inflated residuals drawn with replacement from the pool of the residuals *for that cell*. Using the bootstrapped response vector, recalculate $\Delta\text{sic}^*$(and whatever other quantities might be of interest). 4) Repeat step 3 a predetermined number ($n_B$) times and store the $n_B$ values of $(\Delta\text{sic})^*$. 5) Produce an EDF and summary statistics from the $\Delta\text{sic}^*$ values (and for whatever other quantities).

Once an estimated distribution function, either parametric or nonparametric, for ΔSIC is available, several other important statistics can be calculated. These include the mean and the median values of the bootstrapped evidence function, the bootstrap confidence intervals and the apparent or approximate reliability (aR) of the model identification. The aR is the proportion of the evidence function distribution that falls above 0. Another interesting statistic is the estimate of $\Delta K$ (Eq. 2)given by

$$\widehat{\Delta K} = \Delta \mathrm{SIC}/n\ . \qquad (47)$$

An EDF for $\widehat{\Delta K}$ values, along with confidence intervals for $\Delta K$, can be informative. Unlike Δsic comparisons, which are thought to be valid only within the same dataset, $\Delta K$ comparisons can be made between different datasets [4], as long as a common base for calculating logarithms has been used.

Additional post-data analyses can revolve around model evaluation. The traditional diagnostics using residuals for normal-based models should be performed with the selected model. Outliers and influential observations should be detected and investigated. The designed error properties of the evidential analysis were constructed under the assumption that one of the two normal linear model distributions is an adequate representation of the probabilistic mechanism generating the data, and the assumption-checking analyses help to bolster confidence in the evidential results.

9. Example: two-factor analysis of variance

A two-factor analysis of variance is a warhorse of agricultural studies, in which the response of growth or yield of the agricultural product is measured in the presence of different levels of two categorical treatment factors (such as nitrogen and phosphorus levels). Interactions between the factors can be estimated when the treatment level combinations have more than one observation. Interest typically lies in which combinations of factor levels produce the best yields. A particular concern is whether the two factors interact, that is, whether levels of one factor affect the effect strengths of the other factor levels.

The comparison of the full linear model containing interactions with the restricted linear model lacking interactions is conveniently accomplished with Formulation 1 of hypothesis testing (Eqs. 19 and 20). If Factor 1 has $l_1$ levels and Factor 2 has $l_2$ levels, then the matrix $\boldsymbol{X}$ (in the "leave one column out" or "means" coding) will have $n$ rows and $l_1 \times l_2$ columns, consisting of a column of 1's for the intercept, $(l_1 - 1)$ columns of indicator variables for the levels of Factor 1 (each column containing 1's and 0's), $(l_2 - 1)$ columns of such indicator variables for the levels of Factor 2, and $(l_1 - 1) \times (l_2 - 1)$ columns of elementwise products of all the Factor 1 and Factor 2 indicator variables representing interactions.

For the analysis of whether interactions are important or not, model 1 has matrix $\boldsymbol{X}_1$ containing just the intercept, the columns of indicator variables for Factor 1, and the columns of indicator variables for Factor 2. The additional matrix $\boldsymbol{X}_2$ in model 2 has the interaction columns. The noncentrality quantity $\lambda$ (Eq. 24) is straightforward to calculate with matrix-based computational software. However, for designing an evidential analysis targeted at the overall interaction effect, one can specify the largest relative effect size $\delta\sigma$ of interactions acceptable for selecting the non-interaction model (model 1) and avoid the need to calculate the $\lambda$ formula, provided the design is simple.

The example data (Table 1) are from Ott and Longnecker ([25]; their example 15.8) and consist of fruit yields from 24 citrus trees. The original study is not cited in the Ott and Longnecker textbook, but the data are iconic of many such problems seen in statistics consulting centers. Factor 1 levels are 3 tree varieties, and Factor 2 levels are 4 pesticide types. The design is balanced with 2 trees for each factor combination. Thus, $n = 24, r = 12, q = 6$. Such a design is equivalent to a one-way analysis of variance with 12 levels. Here for illustration we will use the $n\delta^2$ formulation for $\lambda$. The standard NP hypothesis test of model 1 (null) versus model 2 (alternative) fails to reject the hypothesis of interactions ($F = 1.80,\ p = .18$). The evidence function has value $\Delta\text{sic} = -3.66$. If the allowable relative effect size for the null model is half of a standard deviation, then one can use $\delta = .5$. The resulting value of the noncentrality quantity becomes $\lambda = n\delta^2 = 6$. Fixing $M_1$ and $M_2$ to be no more than $.05$, the $k$ thresholds become roughly $k_1 = -12.9$, $k_2 = 13.3$. We have $k_1 < \Delta\text{sic} < k_2$, indicating insufficient evidence to favor either model. However, if $\delta = 1$, then $k_1 = -2.16$, and the data provide strong evidence that the relative effect of interactions is no larger than one standard deviation. The value of $\delta$ for which $P_2 = 0.05$ is around $\delta \approx .94$.

Table 1. Evidential analysis of a two-way factorial study: is there substantial interaction between the factors? The observations of the response variable are fruit yields of 3 varieties of citrus trees (8 trees of each variety, 24 trees total). Two trees of each variety are allocated to treatment with one of four pesticides. In the main effects only model (model 1), the matrix $\boldsymbol{X}_1$ has 6 columns: intercept column (all ones), 2 indicator columns (ones and zeros) designating 2 of the tree varieties, 3 indicator columns (ones and zeros) designating 3 of the pesticide types). In the full model with interactions (model 2), the additional matrix $\boldsymbol{X}_2$ has 6 columns of elementwise products of each pair of indicator variables from $\boldsymbol{X}_1$. Thus, $n = 24, r = 12, q = 6$. We assume that the vector $\boldsymbol{y}$ of observations arises from a multivariate normal($\boldsymbol{X\beta}, \sigma^2\boldsymbol{I}$) distribution, where $\boldsymbol{X\beta} = \boldsymbol{X}_1\boldsymbol{\beta}_1 + \boldsymbol{X}_2\boldsymbol{\beta}_2$. Model 1 will be acceptable provided evidence strongly suggests that $\lambda < n\delta^2$ (relative effect size of interaction is acceptably small), and model 2 will be favored if evidence strongly suggests otherwise. Results are given for values of $\delta$ at $.5$ and at 1. The maximum probabilities of misleading evidence were set at $\gamma_1 = \gamma_2 = .05$. Results for each value of $\delta$ include $f$, $\lambda = n\delta^2$ (noncentrality parameter for the noncentral F distribution calculated on the boundary between the models), $\psi_1, \psi_2$ (respectively the $\gamma_2$th and $(1 - \gamma_1)$th quantiles of the noncentral F($q, n - r, \lambda$) distribution), $k_1, k_2$ (the lower and upper threshold values indicating strong evidence for model 1 and model 2 respectively, with the interval in between indicating inconclusive evidence), and Δsic (the value of the evidence function). We find that $k_1 <$ Δsic $< k_2$ when $\delta = .5$, indicating insufficient evidence favoring either model, while Δsic $< k_1$ when $\delta = .1$, indicating strong evidence for a relative effect size of interactions less than one standard deviation. The smallest relative effect size for which there would be strong evidence is $\delta = .94$.

| Tree Variety | Pesticide type 1 | 2 | 3 | 4 |
|---|---|---|---|---|
| 1 | 49, 39 | 50, 55 | 43, 38 | 53, 48 |
| 2 | 55, 41 | 67, 58 | 53, 42 | 85, 73 |
| 3 | 66, 68 | 85, 92 | 69, 62 | 85, 99 |

$n = 24$, $r = 12$, $q = 6$

$f = 1.80$, $p = .18$

$\Delta\text{sic} = -3.66$

$\delta = .5$, $\lambda = 6$, $P_2 = .45$

$\psi_1 = .584$, $\psi_2 = 5.69$

$k_1 = -12.9$, $k_2 = 13.3$

$\delta = 1$, $\lambda = 24$, $P_2 = .03$

$\psi_1 = 2.05$, $\psi_2 = 12.9$

$k_1 = -2.16$, $k_2 = 29.2$

$\delta = .94$, $\lambda = 21.2$, $P_2 = .05$

The bootstrap EDF of ΔSIC reveals a heavy-tailed distribution (Figure 2). The parametric and nonparametric bootstrap versions were each based on 1024 bootstrap samples. The $.05$ and $.95$ quantiles are respectively about $-10$ and $+20$, indicating a wide variability of ΔSIC values is to be expected for the sample sizes in the citrus study. The means of the bootstrapped ΔSIC values of 2.7 (parametric) and 3.5 (nonparametric) indicate weak evidence for model 2. The aR is estimated as 0.61 by using a parametric bootstrap and 0.62 by the non-parametric bootstrap also supports very equivocal evidence for interactions. The EDFs reinforce the conclusion that the data were insufficient to resolve the interaction question more sharply.

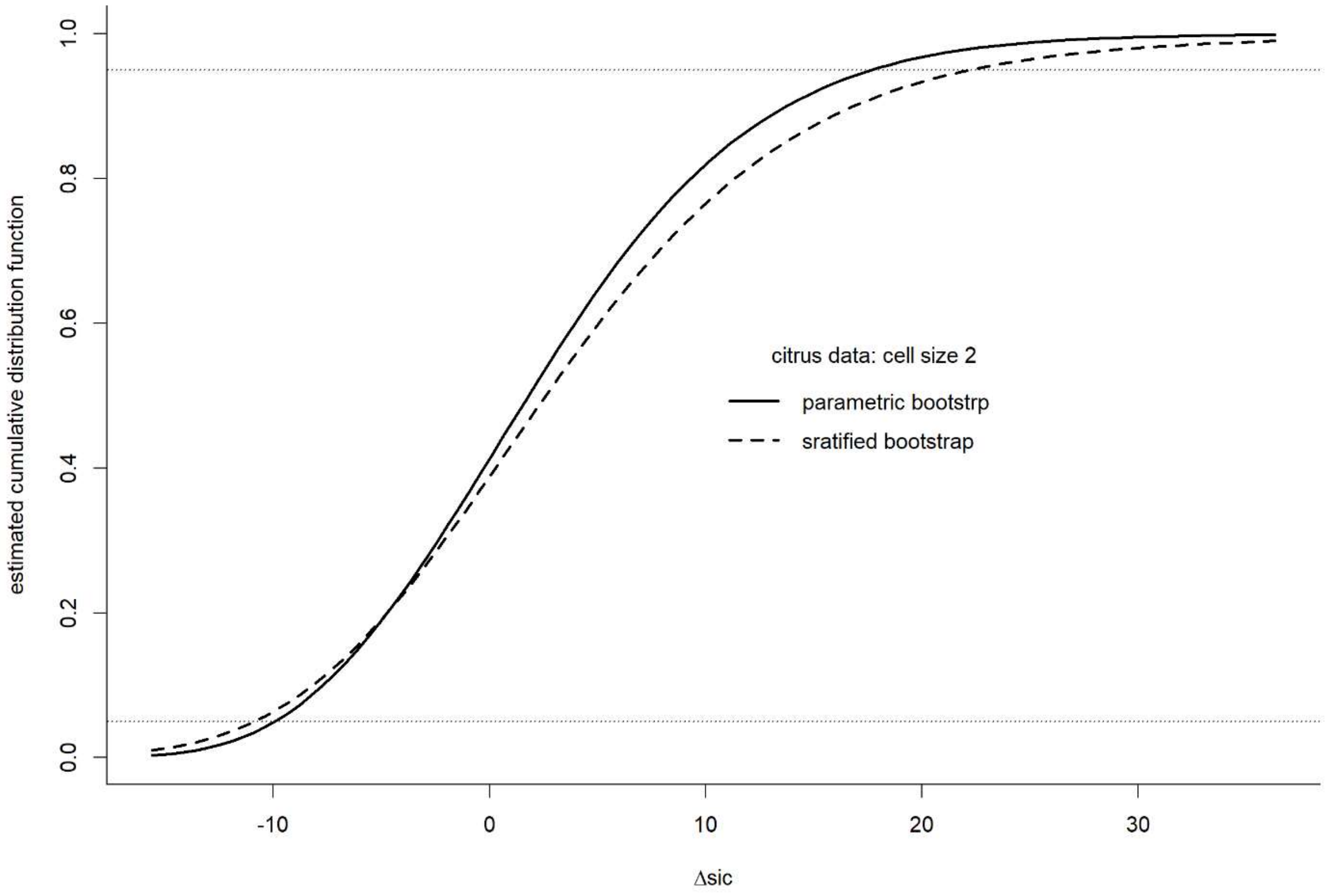

Figure 2. Curves: estimated cdf of ΔSIC for the citrus tree example (two-factor analysis of variance, Table 1, with model 1 representing no interactions, model 2 representing interactions) using parametric (solid) and nonparametric (dashed) bootstrap with 1024 bootstrap samples. Dotted horizontal lines depict .05 and .95 levels.

Simulation can be used to study the effect of larger sample sizes. In Figure 3, 90% confidence intervals for $\Delta K$ are depicted when observations are added to each cell (treatment combination). The data are simulated from the estimated model 2 (representing interactions). For each hypothetical sample size, confidence intervals for $\Delta K$ were generated with 1024 parametric and nonparametric bootstraps for the distribution of the estimate of $\Delta K$ given by Eq. 47.

To depict the expected behavior of such intervals the confidence points (0.05, 0.50, 0.95) from 1024 simulated data sets are averaged in Figure 3. The solid horizontal line indicates equal evidence for model 1 and model 2. The dotted horizontal line indicates the pseudo-true difference of KL divergences for the citrus model used in the simulations. For cell sizes of 4 and above, the parametric and nonparametric intervals are almost identical. Note how the evidence function based on the SIC approaches truth from below, reflecting the complexity-averse nature of the SIC.

The nonparametric confidence interval for $\Delta\text{SIC}$ in Figure 2 was roughly 20% greater than the parametric interval. The same pattern is seen in Figure 3 for the per-cell sample size ($n_i = 2$) in the simulations, but the difference of parametric and nonparametric intervals rapidly dimishes as sample size increases. The result suggests that only 2 observations per cell might be fundamentally uninformative about interactions, under the prevailing level of noise in the data.

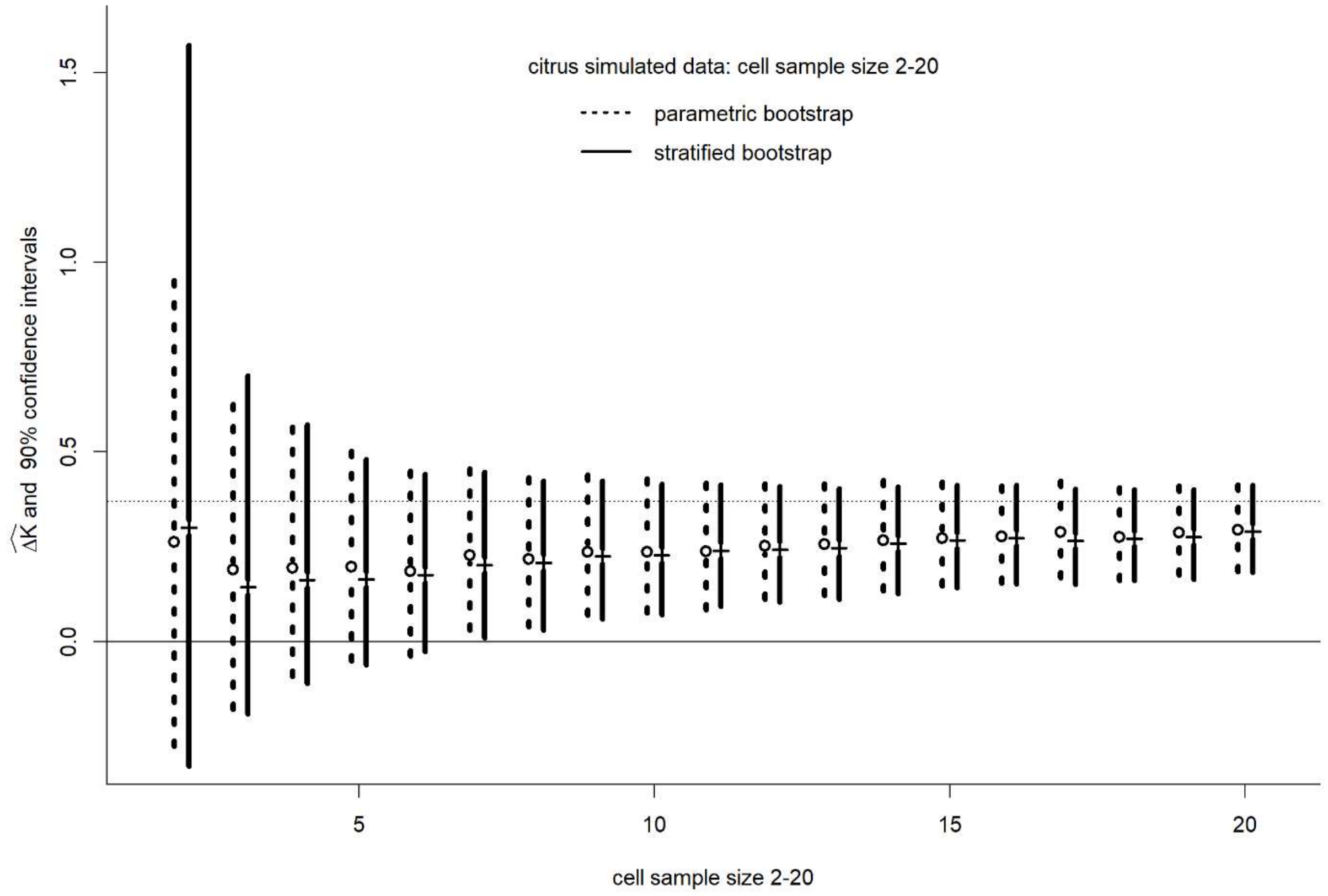

Figure 3. The effect of sample size on the uncertainty of an evidential estimation. The data are simulated from the estimated model 2 (representing interactions). For each data set, confidence intervals were generated with 1024 bootstraps. To depict the expected behavior of such intervals the confidence points (0.05, 0.50, 0.95) from 1024 simulated data sets are averaged. The solid horizontal line indicates equal evidence for model 1 and model 2. The dotted horizontal line indicates the pseudo-true difference of Kullback-Leibler divergences in the simulations.

## 10. Discussion

The drawbacks of NP testing have been widely discussed in the literatures of the sciences and social sciences. In particular, from the standpoint of model selection, the "null model is always false" in NP testing, and the testing results are unhelpful toward evaluating the amount of compromise involved in using one model over the other. The main problem is that the rigid behavioral threshold of $\alpha = 0.05$ (or whatever desired test size) in NP testing, along with the only slightly more nuanced reference to the $p$-value itself, make limited use of the information in the data. A test fails to reject the null hypothesis: what does that really mean from the standpoint of study objectives? Confidence intervals, as ranges of parameter values for which NP tests fail to reject the null model, are more informative, but the $0.05$ (or whatever) confidence level remains as an impediment to interpreting the consequences of using either model.

Inferences from the evidential approach differ fundamentally from those of the NP approach. The standard NP setup for analyzing nested linear models bases the error rate for the decision threshold on the central F distribution (or central chisquare in other model families). The central F distribution is based literally on the parameter constraints in the null hypothesis: the parameter (or parameter vector) equals zero (or whatever) to all decimal places. Somewhat concealed in the NP setup is that the null hypothesis is never true; the implicit parameter constraints in the null hypothesis are instead a "practical set of measure zero." The real concern almost always revolves around whether the difference of a parameter from zero is of practical importance. However, this concern only surfaces in the NP setup when questions are asked about how big an effect (departure from zero) could be detected with a proposed study design or about how severe a test has the null hypothesis survived.

Evidential analysis departs from NP testing by specifying up front how much of a departure from a parameter constraint is of importance to the investigation. The focus shifts from Type 1 and Type 2 error probabilities to probabilities $M_1$ and $M_2$ of misleading evidence. The specification of a parameter zone of negligible effect for model 1 then leads then to the use of the noncental F distribution for setting up error rates and decision thresholds. The noncentral F distribution is heavier-tailed, having a variance greater than that of the corresponding central F distribution (Figure 1). The resulting inferences, although often more sobering and nuanced, are of more practical value for building more useful models.

Valuable information is provided by evidential analysis over and beyond NP testing. Evidential analysis, by contrast to NP testing, provides an assessment of how large a departure from the null hypothesis (model 1), in comparison to the general noise level of the data, can be ruled in or ruled out. As well, evidential analysis provides a more complete understanding of the uncertainty accompanying the results. The two-way ANOVA example analyzed above illustrated a typical problem arising in day-to-day experimental science: an effect, here an interaction of factors, has magnitude just off the radar in ordinary NP testing. A $p$-value of .18 in the NP test for interaction is scientific pablum, in that a mild effect of unknown magnitude and unknown import might or might not be present. Should more data be collected? Should the estimated alternative model be reported and used? What is lost by using the estimated null model? In the evidential approach, the comparison of the magnitude of the per observation effect with the per observation standard deviation helps address these questions. The investigation can focus on how large an effect is acceptable to be lost in noise, as the evidential analysis provides an idea of how small an effect is warranted by the data. In addition, as represented by the bootstrap EDFs for the evidence function, the analysis provides a clearer idea of the scale of uncertainty present in the conclusions.

One will typically find that larger sample sizes are indicated when applying evidence concepts to ordinary NP testing situations. This finding is not illusory. The evidential framework for normal linear models is based on noncentral distributions; the calculations are similar to power calculations in experimental design, which in most consulting statisticians' experiences have not ever made any investigators happy. Low powers, such as .6, are often used as design benchmarks for NP tests, and the result is higher uncertainty about the conclusions and about replicability. The statistical distributions of evidential quantities have heavy tails (Figs. 2, 3), and to obtain sharp conclusions it is not uncommon for evidential design to call for sample sizes to be increased by an order of magnitude, although in our example, a factor of 4 might have brought the lower end of a 90% confidence interval for $\Delta K$ above 0 (Figure 3).

The conclusions of NP testing depend, sometimes sensitively, on the assumption of correct model specification. The null hypothesis in NP testing, formed for instance by zeroing-out one or more parameters, is seldom strictly correct, but parameters representing an effect size negligible for practical purposes might belong to the model closest to the data-generating model. With fixed test size $\alpha$, an NP analysis will asymptotically reject such an acceptable model. In an evidential analysis, there is reason to retain some confidence in the results even in the presence of moderate violations of assumptions. If the misleading evidence probabilities $M_2$ and $M_1$ are redefined as the probabilities of picking the model farthest from truth $g$ (in the KL divergence sense), then both probabilities go toward zero as sample size increases, once the values of $k_1$ and $k_2$ are set [7]. Thus, evidential analysis retains a robustness of sorts to model misspecification, although if models are misspecified the actual values of $M_2$ and $M_1$ would be unknown.

Moreover, a central tenant in modern evidential statistics, at least that branch stemming from Lele [4], is the idea that some degree of model misspecification is ubiquitous. The foundational evidential concepts assume that neither model 1 nor model 2 generated the data but rather the data came from an unknown model with pdf $g(y)$ [4, 7, 8]. The popularity of parametric statistical models has persisted long beyond the advent of bootstrapping in the late 1970s, due to the insights such models can contribute to the structure of phenomena. However, the uncertainty of conclusions in parametric modeling is often underestimated in the parametric framework. Thus, whenever possible, assessing uncertainty by estimating $g(y)$ directly, along with distributions of statistics for comparing model 1 with model 2, using nonparametric bootstrapping seems a compelling plan [8]. There are promising avenues toward estimating the distribution of $\Delta\mathrm{SIC}$ through nonparametric bootstrap estimation of $g$ [8]. We see in Figure 3 that under the correct model assumption a particular stratified nonparametric bootstrap can recapitulate parametric confidence intervals at astonishingly small cell sizes (echoing the conclusion of earlier work on the stratified bootstrap [22]). Thus, an analyst can comfortably use a stratified bootstrap to add another level of protection from the effects of misspecification.

Because of the misspecification risk, the importance of model evaluation with post-analysis diagnostics in NP testing has been stressed widely. The validity, or lack thereof, of the NP conclusions in any particular situation can engender much concern. An evidential approach, by contrast, can pre-specify a zone of null hypothesis tolerance as well as explore the strengths of evidence for other parameter zones. Model evaluation diagnostics in an evidential analysis can thereby be focused more on the usefulness, or lack thereof, of the candidate models.

Although we have treated here an evidential approach to the classical NP setup of two models, one nested within the other, the evidential approach extends naturally to models that are merely overlapping and even to models that are nonoverlapping [7, 8]. Such scenarios would be encountered, for instance, when selecting among many predictor variables for inclusion in a multiple regression. The methods of using NP testing for variable selection such as stepwise regression always seemed contrived but were the only solutions available before the widespread adoption of model selection indexes. In the evidential approach, such indexes are turned into evidence functions with probability distributions and associated inferences of uncertainty. However, the distribution theory for evidence functions in such scenarios is asymptotic (e.g. [27]), and implementations will generally rely on bootstrapping and simulations [8].

Other families of non-normal statistical models (multinomial, Poisson, etc.) with fixed effect covariates fall under the broad umbrella of "generalized linear models" [28]. For such models, symbolic estimates of ML estimates are usually not available and numerical optimization of log-likelihood functions is employed. For NP testing (and associated confidence intervals), exact forms of distributions for test statistics are not available, and testing relies on asymptotic approximations. The asymptotic distribution of the generalized likelihood ratio statistic (Eq. 6) under the null hypothesis (under regularity conditions) is well-known to be a chisquare distribution [19] and is the standard source of $p$-values printed by software packages. For addressing such model comparisons with an evidential framework, the distributions of the test statistic under the alternative models will be central to the inferences. The asymptotic distribution of the generalized likelihood statistic under alternative models (in a mathematically localized sense) is a noncentral chisquare distribution [29, 30, 31]. The forms of the KL divergences, noncentrality quantities, and the adequacy of asymptotic approximations differ among model families. A future paper by the authors will explore many of the issues involved.

Supplementary Materials: R functions needed to conduct the analyses in this paper, principally lm.rEV, can be found at <https://urldefense.com/v3/__https://github.com/jmponciano/EvidentialAnalysis__;!!JYXjzlvb!k1FhA5s_M2md2MUfoygLFDzILH4h3Q51-cziSSe0pB9AYyJYeJa-KT8WbQrLHOtDi0I8EIJe3fg51tYiGA$>

Funding: This research received no external funding.

Institutional Review Board Statement: Not applicable.

Informed Consent Statement: Not applicable.

Data Availability Statement: Data appear in Table 1.

Conflicts of Interest: The authors declare no conflicts of interest.